\documentclass[prl,twocolumn,showpacs]{revtex4} 
\input{psfig.sty} 
\begin{document} 
 
\title{ 
  Hierarchical Chain Model of Spider Capture Silk Elasticity 
} 
 
\author{Haijun Zhou$^1$ and Yang Zhang$^2$} 
\affiliation{ 
  $^1$Max-Planck-Institute of Colloids and Interfaces, 14424 Potsdam, Germany \\
  $^2$Center of Excellence in Bioinformatics, University at Buffalo, 
  901 Washington St., Buffalo, NY 14203 
} 
 
\date{21 January, 2005} 
 
\begin{abstract} 
  Spider capture silk is a biomaterial with both high strength and high 
  elasticity, but the structural design principle underlying these 
  remarkable properties is still  unknown.  
  It was revealed recently by atomic force microscopy that
  an exponential force-extension relationship holds both for 
  capture silk mesostructures and for intact capture silk fibers 
  [N.~Becker {\em et al.}, Nat.~Mater.~{\bf 2}, 278 (2003)]. 
  In this Letter a simple hierarchical chain model was proposed
  to understand and reproduce this striking observation.  
  In the hierarchical chain model, a polymer is composed of many 
  structural motifs which  organize into structural  modules and 
  supramodules in a 
  hierarchical manner. Each module in this hierarchy has its own 
  characteristic force. The repetitive patterns in the 
  amino acid sequence of the major flagelliform protein of spider
  capture silk is in support of this model. 
 
 [Phys. Rev. Lett. {\bf 94}, 028104 (2005)]
\end{abstract} 
 
 
\maketitle 
 
The capture silk is a natural material produced  by orb-web weaving spiders.  
Spiders rely on it to entrap flying preys  
\cite{VollrathF1992}. Like the spider dragline silk and  
many other naturally occurring silks,  
the capture silk has a tensile strength that is 
comparable to steel; but, unlike steel, it is also extremely elastic, with 
the ability to be stretched to almost $10$ times its relaxed contour length 
without breaking \cite{GoslineJ1999,BeckerN2003}. This perfect combination 
of strength and extensibility conveys a high degree of toughness to 
the capture silk: its breakage energy per unit weight 
is more than $20$ times that of a high-tensile steel 
\cite{GoslineJ1999}. With the aim to produce 
synthetic silks with similar mechanical properties, 
materials scientists have devoted  
many experimental and computational 
efforts to the understanding of spider silk's structural organization 
\cite{KaplanD1994}. 
Despite these painstaking efforts, 
the mechanism behind spider silk's remarkable strength and elasticity 
is still largely missing, partly because of the difficulty to obtain  
high-quality crystallized structures of silk proteins. 
 
Single-molecule manipulation methods have been recently applied on 
spider silks to obtain very detailed information on 
their force-extension response 
\cite{BeckerN2003,OroudjevE2002}.  
In a recent experiment, 
Hansma and co-workers \cite{BeckerN2003} attached  
capture silk mesostructures (probably composed of
a single protein molecule)  or intact capture silk fibers to  
an atomic force microscopy tip and 
recorded the response of the samples to external stretching force.  
They found a remarkable exponential relationship between the 
extension $x$ and the external force $f$,
\begin{equation} 
  \label{eq:eq01} 
  f\propto  \exp( x/\ell ) , 
\end{equation} 
where $\ell$, the length constant referred to in \cite{BeckerN2003},
is a fitting parameter whose physical meaning needs to be
decided (see below).
$\ell= 110 \pm 30$ nm for a capture silk molecule and
$\ell = 11 \pm 3$ mm for an intact capture silk fiber \cite{BeckerN2003}.  
The length constant of a silk fiber is about $10^5$ times that 
of a silk molecule; its relaxed contour length is also about $10^5$
times that of a molecule \cite{BeckerN2003}. 
It appears that $\ell$ holds an approximately 
linear relationship with the contour length.
 
The exponential force-extension curve is significantly different from  
the data observed during stretching single double-stranded  (ds)
DNA molecules \cite{SmithS1992} 
or single titin proteins 
\cite{RiefM1997,KellermayerM1997}. 
The behavior of dsDNA can be understood by 
the wormlike chain model of entropic elasticity \cite{MarkoJ1995,ZhouH1999}, 
and that of titin by a two-level system coupled with entropic elasticity 
\cite{RiefM1998}. 
Similar exponential force-extension data were also observed 
by Dessinges {\it et al.} when they stretched a single-stranded (ss) DNA  
molecule at low salt conditions 
\cite{DessingesM2002}. 
The data was explained as a result of the interplay of electrostatic 
repulsion and entropic elasticity \cite{ZhangY2001,DessingesM2002}. 
However, the success of this model depends on the specific ionic 
concentration (i.e., low salt and high electrostatic interaction); 
and it could not naturally reproduce the exponential  
stretching data at extremely high force  
(i.e., when the extension is larger than 
$1.1$ times  the ssDNA backbone length) if  
higher order deformation energy terms are not  
included \cite{DessingesM2002}. 
In  the spider capture silk experiment, the exponential behavior 
was observed at both fluid and air within a force range 
from about $10^0$ piconewton (pN) to about $10^6$  pN \cite{BeckerN2003}.

Equation~(\ref{eq:eq01}) indicates the following: ($i$) 
Because the capture silk is highly extensible, 
a great amount of extra length must have been stored in its relaxed form. 
($ii$) Since extension increases with force logarithmically, 
some fraction of the stored length must be  easy to be pulled out, 
some fraction be harder to be pulled out, and till some other 
fraction be even harder to be pulled out.  
To model this kind of heuristic cascading responses, here we 
propose  a {\em hierarchical chain model} 
 for spider capture silk. 
In the hierarchical chain model, the polymer is composed of many basic  
structural motifs; these motifs are then organized into a hierarchy, 
forming structural 
modules on more and more longer length scales.  
At the deepest hierarchy level  $h_{m}$, the structural motifs could 
be $\beta$-sheets,  
$\beta$-spirals, helices \cite{KaplanD1994} or  microcrystal 
structures \cite{VollrathF2001}.  
The interactions among some of these motifs are  much  
more stronger than their interactions with other motifs,  
therefore they form a structural module at the hierarchy level 
$(h_{m}{-}1)$. 
These level-$(h_{m}{-}1)$ modules are then merged into 
level-$(h_{m}{-}2)$ modules through their mutual interactions. 
This merging process is continued;  
and  finally at the global scale, the whole spider silk string 
is regarded as a single module of  the hierarchy level 
$h=0$. 
 
In nature, the structures of many biomaterials are indeed 
hierarchically organized. 
As a composite material, chromosome is a mixture of DNA, histone, and other 
non-histone structural proteins \cite{vanHoldeK1989}.  
The DNA molecule first wraps 
onto histone proteins to form nucleosome particles,  
the basic units of chromosome. 
With the help of H1 histone, this  
linear sequence of nucleosomes are then coiled and folded to form the 
$30$-nm chromatin fiber. With the help of other scaffold proteins, the
chromatin 
fiber is then further coiled and folded at several levels to form the compact 
chromosome structure.
As another example, the amino acid sequence of a protein first 
forms basic structural modules of $\alpha$-helix and $\beta$-sheet,  
called secondary 
structures. By different ways of connections of these secondary 
modules, the protein constructs various tertiary topologies that are 
critical for its specific biological functions. 
In a higher level, these tertiary domains then form  
quaternary structures of protein complexes consisting of multiple 
chains \cite{BrandenC1999}.  
Recent experimental and theoretical studies have shown that the 
force--induced unfolding of protein molecules is indeed processed in  
a hierarchical way, {i.e.} the tertiary structure precedes the secondary 
structures to be pulled over at increasing external forces 
\cite{RiefM1997,LuH1998}. Our hierarchical chain 
model may also serve as a framework to understand the mechanical 
property of these biomaterial systems.

\begin{figure}[t] 
  \hspace*{0.8cm}\psfig{file=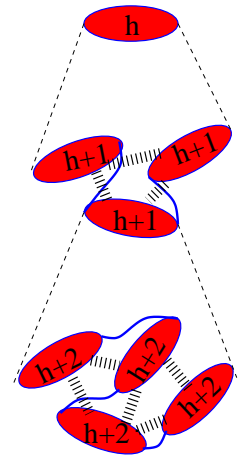,height=6.0cm} 
  \caption{ 
    \label{fig:fig01} 
    The hierarchical chain model. At each hierarchy level $h$ 
    a structural module $M_h$ is composed of a tandem sequence 
    of $m_h$ submodules $M_{h+1}$ of hierarchy level $h+1$.  
    Under external stretching, $M_h$ responds by ($i$)  
    adjusting the arrangements of 
    its  $m_h$ subunits and making them  more aligned along the 
    force direction, and ($ii$) extending these $m_h$ subunits. 
    The thick broken lines between submodules of each hierarchy level 
    indicate the existence of sacrificial bonds. 
  } 
\end{figure} 

Consider a polymer  of contour length $L_0$. It is regarded as the module 
of hierarchy level $h=0$ (the module $M_0$). 
$M_0$ is composed of a tandem sequence of $m_0$ subunits $M_1$
of contour length $L_1=L_0/m_0$ (Fig.~\ref{fig:fig01}).  
Under the action of an external force field $f$, the extension of 
$M_0$ is denoted as $x_0(f)$.  It was observed that water-induced 
protein mobility is a significant contribution to capture silk elasticity 
\cite{BonthroneK1992}. Therefore, we decompose  $x_0$ into two parts: ($i$)  
the extension $\Delta x_0$ 
caused by the displacement and position-rearrangement of these $m_0$ 
subunits and, ($ii$) the total extension $m_0 x_1$ caused by the 
inherent deformations of these $m_0$ subunits: 
\begin{equation} 
  \label{eq:eq02} 
  x_0(f)=\Delta x_0(f) + m_0 x_1(f). 
\end{equation} 
Similarly, at the hierarchy level $h=1$, each unit is itself composed of  
$m_1$ level-$2$ subunits $M_2$ of length $L_2=L_0/(m_0 m_1)$. Therefore,  
$x_1(f)$ can be written as  
$x_1(f)=\Delta x_1(f) + m_1 x_2(f)$, 
where $\Delta x_1(f)$ is the contribution to the extension of a level-$1$ module 
due to the displacement and position-rearrangement of its $m_1$ subunits; 
and $x_2(f)$ denotes the extension because of the inherent deformation of a 
level-$2$ subunit $M_2$.  
As this structural hierarchy is continued, we arrive at the  
following expression concerning the total extension $x(f)$:
\begin{equation} 
  x(f)= \Delta x_0(f)+ \sum_{h=0} m_0 m_1 \ldots m_h  \Delta x_{h+1}(f). 
  \label{eq:eq04} 
\end{equation}

It is reasonable that the average extension $\Delta x_h(f)$ of a 
$M_h$ contributed by the reorientation or rearrangement of its level-$(h+1)$ 
subunits is proportional to the relaxed contour length $L_h$ of this 
module. To facilitate the following analytical calculation,  
let us assume $\Delta x_h(f)$ has the following  
non-linear form (we will show that the final force-extension 
relationship is not sensitive to the specific 
assumption made here): 
\begin{equation} 
  \label{eq:eq05} 
  \Delta x_h(f)=\left\{ 
    \begin{array}{ll} 
      \alpha L_h f / f_h,\;\; & f< f_h\\ 
      \alpha  L_h, & f\geq f_h 
    \end{array} 
  \right. 
\end{equation} 
where $\alpha$ is a dimensionless proportional constant; 
$f_h$ is the characteristic force needed to displace and 
rearrange the positions of the $m_h$ submodules contained in a $M_h$
(during this process some sacrificial bonds are broken).  
Equation~(\ref{eq:eq05}) is in agreement with the 
experimental observation  \cite{BeckerN2003}  
that, between adjacent rapture 
events,  a capture silk responds to external stretching in a 
linear way.  
Denote $\Delta E_{h}$ as the energy cost of  breaking all the sacrificial 
bonds between  a level-$h$ module's  $m_h$ subunits. 
From Eq.~(\ref{eq:eq05}) we know that
$f_{h+1} / f_{h}= m_h \Delta E_{h+1} / \Delta E_{h}$. 
Consider a level-(h+2) module $M_{h+2}^a$, it is in $M_{h+1}^a$ which
in turn is in $M_{h}^a$. $M_{h+2}^a$ feels an {\em internal} energy
$\epsilon$ due to its interaction with other subunits in $M_{h+1}^a$, and it
feels an {\em external} energy $\epsilon^\prime$ due to its interaction with other
subunits in $M_{h}^a$ but not in $M_{h+1}^a$.  
Based on Fig.~\ref{fig:fig01}, we know that $\Delta E_{h+1}= m_{h+1} \epsilon /2$ and
$\Delta E_{h} = m_h m_{h+1} \epsilon^\prime /2$.
The hierarchical organization of the polymer requires that $\epsilon > \epsilon^\prime$, so as to
ensure that structural modules of shorter length scales will be 
formed earlier.  Based on these considerations,
we arrive at the  following self-similar scaling form: 
\begin{equation} 
  \label{eq:eq06} 
  f_{h+1} = (\epsilon / \epsilon^\prime ) f_h = \beta f_h \;\;\;\;\;\;(\beta \equiv \epsilon/ \epsilon^\prime  > 1).
\end{equation} 
The parameter $\beta$ characterizes the
degree of coherence in the modular organization of the polymer:
a large $\beta$ value means that a submodule has much stronger internal
interactions compared with its external interactions.

From Eqs.~(\ref{eq:eq04}) and (\ref{eq:eq05}) we find that when 
$f_{h-1} < f \leq f_{h}$  
\begin{equation} 
  \label{eq:eq07} 
  {{\rm d} x(f) \over {\rm d} f} = \sum\limits_{h^\prime = h} {\alpha L_0 \over f_{h^\prime} } 
  = {\alpha \beta L_0 \over \beta -1} f^{-1}. 
\end{equation} 
Equation~(\ref{eq:eq07}) therefore recovers the experimental 
exponential force-extension relationship of Eq.~(\ref{eq:eq01}) with 
\begin{equation} 
  \label{eq:eq08} 
  \ell={\alpha \beta \over \beta -1} L_0. 
\end{equation} 
The length constant $\ell$ is proportional to the relaxed contour length 
$L_0$ of the whole polymer, 
in consistence with Ref.~\cite{BeckerN2003}.  
 
\begin{figure}[b] 
  \psfig{file=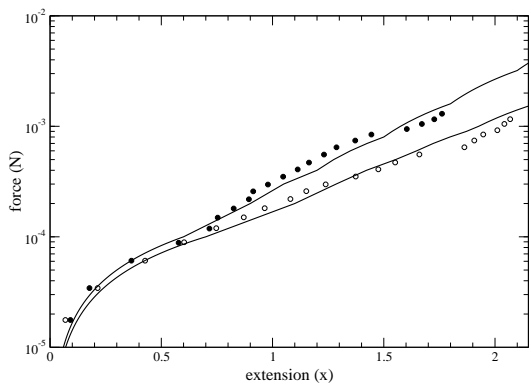,width=9.0cm,angle=270} 
  \caption{ 
    \label{fig:fig02} 
    Exponential force-extension relationship for the hierarchical chain model. 
    Equation~(\ref{eq:eq05}) is used in the numerical calculation. The 
    parameters are $f_0=10^{-4}$ N, $\alpha=0.3$, and $\beta=2$ 
    (the upper curve) or $\beta=1.75$ (the lower curve). 
    Extension is in units of $L_0$. Symbols are  experimental data from
    Fig.~4 of \cite{BeckerN2003}.
  } 
\end{figure} 
 
Figure \ref{fig:fig02} demonstrates the numerically calculated 
force-extension curve  based on Eqs.~(\ref{eq:eq04}), (\ref{eq:eq05}), 
and (\ref{eq:eq06}). As a comparison, the experimental data \cite{BeckerN2003}
on intact spider capture silk is also shown. 
As $\beta \simeq 2$ and the exprimental exponential force range 
is roughly from $6\times 10^{-5} N$ to
$10^{-3} N$, it appears that
$4$--$5$ levels of hierarchy were probed.
 
\begin{figure}[b] 
  \psfig{file=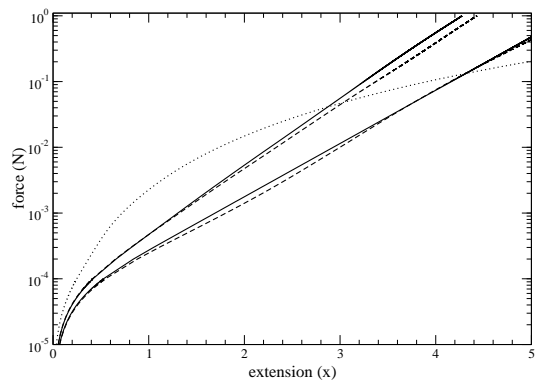,width=9.0cm,angle=270} 
  \caption{ 
    \label{fig:fig03} 
    The force-extension relationship of a hierarchical chain is insensitive 
    to the assumption made to the response $\Delta x_h(f)$ of Eq.~(\ref{eq:eq05}),  
    but is sensitive to the hierarchical scaling form of the characteristic  
    force $f_h$.  
    The solid lines are obtained by assuming $\Delta x_h(f)$ has the form of  
    Eq.~(\ref{eq:eq09}), while other parameters are the same as those in  
    Fig.~\ref{fig:fig02}.  
    The dotted line shows the change in the force-extension curve when 
    additionally a power-law form of $f_h =f_0 h^{\gamma}$ with  
    $\gamma=3.0$ is assumed for 
    $f_h$. The dashed lines are obtained by assuming 
    Eq.~(\ref{eq:eq09}) and Eq.~(\ref{eq:eq06}), with $\beta$ fluctuating  
    uniformly within $[1.25,2.25]$ (the 
    lower curve) and within $[1.5, 2.5]$ (the upper curve). 
  }  
\end{figure}

The exponential relationship shown in the figure is insensitive 
to our particular assumption Eq.~(\ref{eq:eq05}), as long as  
the elastic response at each hierarchy level is non-linear and bounded.   
As an example, 
the solid curves in Fig.~\ref{fig:fig03} 
show  the resulting force-extension relationship 
when Eq.~(\ref{eq:eq05}) is replaced by  
\begin{equation} 
  \label{eq:eq09} 
  \Delta x_h(f)={\alpha L_h} \bigl( 
  1-\exp(-f/f_h)\bigr). 
\end{equation} 
The same exponential behavior as in Fig.~\ref{fig:fig02}  is obtained.
However, the hierarchical scaling form Eq.~(\ref{eq:eq06}) is needed  
for the exponential force-extension correlations. For example, 
when Eq.~(\ref{eq:eq06}) is replaced by a power-law,  
$f_h \propto f_0 h^\gamma$, the 
response is not exponential (the dotted line in Fig.~\ref{fig:fig03}).  
We also noticed that,  
when in Eq.~(\ref{eq:eq06}) the parameter $\beta$ is not a constant but 
fluctuates over some finite range of $\beta>1$, the resulting force-extension 
curve is still exponential (Fig.~\ref{fig:fig03}, dashed lines). 
 
In summary, we have developed a hierarchical chain model to  
understand the strength and elasticity of spider silks. 
Remarkably, this simple model was able to reproduce 
the peculiar exponential force-extension response of spider capture silk  
reported by Becker {\em et al.} \cite{BeckerN2003}. The model can also 
be used as a framework to understand the elasticity of other spider silks and 
other biopolymers with hierarchically organized structures. 

Becker {\em et al.} \cite{BeckerN2003} have proposed an alternate and 
interesting idea to  
model the spider silk as a branched network of interconnected springs.  
In their model, the system 
responds to external stretching by first unfolding the single spring at the 
root level, then the $m$ springs at the first branching level, then 
the $m^2$ springs at the second branching level, and so on. 
The hierarchical chain model developed here is different  
from Becker {\em et  al.}'s model.  
First, the molecule in the hierarchical chain model consists of a tandem 
sequence of structural modules of different length scales. Therefore, no 
assumption of branching structure is made in our model. 
Second, the response of the chain to external perturbations 
is in a hierarchical manner. If the external force is small,  
only those structural units of length scale comparable to the whole polymer 
length will be displaced and rearranged; structural units at short and moderate 
length scales will remain intact. As the external perturbation is 
elevated, additional structural units at more shorter length scales are also 
deformed. Through such a hierarchical organization, a single polymer chain 
can respond to  a great variety of external conditions; at the same time, it is 
able to  keep its degree of structural integrity as high as possible.  
This hierarchical modular structure 
also indicates a broad spectrum of relaxation times. The modules at the shorter
length scales will have much shorter relaxation times and will be refolded first 
when the external force decreases. This gap in relaxation times 
ensures that, after extension, the spider capture silk will 
return to its relaxed state gradually and slowly. 
This is a desirable feature for 
spider capture silk, because a too rapid contract following the insect's 
impact would propel the victim away from the web. 
 
The simple hierarchical chain model, while appealing, needs further experimental 
validation. This model seems to be supported by recent genetic sequencing  
efforts.  By analyzing the cDNA sequence of the major protein of  
spider capture silk,  the flagelliform protein, it was revealed that the amino-acid 
sequence of  flagelliform  has a hierarchy of modularity  
\cite{GueretteP1996,HayashiC1998,HayashiC2000}. At the basic level, the flagelliform 
sequence is consisted of three repetitive modules (motifs): ($i$) the GPGGX 
motif (Gly-Pro-Gly-Gly-X, X $\in \{$Ala,Ser,Tyr,Val$\}$);  
($ii$) the GGX (Gly-Gly-X, X $\in \{$Ala,Ser,The$\}$); ($iii$) the highly conserved 
spacer motif of length $28$ amino-acids. At the next level, an ensemble 
motif is formed, which is a tandem array of $43{-}63$ GPGGX  followed 
by $6{-}12$ GGX, the spacer and another GGX \cite{HayashiC1998}. 
At the even higher level, the ensemble motif then repeats itself about 
$14$ times to form the flagelliform monomer. The 
variable residues X are not randomly 
distributed along the protein sequence \cite{HayashiC1998}, which may 
encode important structural information. 
The structures of spider capture silks (and other spider silks) therefore  
have the potential to be hierarchically organized. 
 
For spider capture silk, one important experiment will be to 
check the validity of Eq.~(\ref{eq:eq06}) by single-molecule 
force spectroscopy. The characteristic forces $f_h$  
may correspond to the forces at the saw-tooth rapture 
events observed by Hansma and co-workers \cite{BeckerN2003}. Although the 
experimental curves of Ref.~\cite{BeckerN2003} is in agreement with  
Eq.~(\ref{eq:eq06}), more systematic investigations are necessary to 
draw a solid conclusion. 
 
\bigskip 
We are grateful to  {H.~Hansma} for communicating her 
experimental results with us prior to publication and for the
experimental data used in Fig.~\ref{fig:fig02}, and 
to {D.~E.~Makarov} for comments on the manuscript. 
Thanks are also due to {R.~Lipowsky}, {Z.~C.~Ou-Yang},  
and {J.~Skolnick} for their support.

\end{document}